\documentclass{elsart}

\usepackage{amssymb}
\usepackage{graphicx}
\usepackage{dcolumn}
\usepackage{bm}
\usepackage{color}
\usepackage{array}
\newcommand{\PreserveBackslash}[1]{\let\temp=\\#1\let\\=\temp}
\newcolumntype{C}[1]{>{\PreserveBackslash\centering}p{#1}}
\newcolumntype{R}[1]{>{\PreserveBackslash\raggedleft}p{#1}}
\newcolumntype{L}[1]{>{\PreserveBackslash\raggedright}p{#1}}

\begin{document}

\begin{frontmatter}

\title{Impact of Heterogeneous Human Activities on Epidemic Spreading}

\author{Zimo Yang},
\author{Ai-Xiang Cui},
\author{Tao Zhou}\ead{zhutou@ustc.edu}

\address{Web Sciences Center, University of Electronic Science and Technology of China, Chengdu 610054, People's Republic of China}

\begin{abstract}

Recent empirical observations suggest a heterogeneous nature of
human activities. The heavy-tailed inter-event time distribution at
population level is well accepted, while whether the individual acts
in a heterogeneous way is still under debate. Motivated by the
impact of temporal heterogeneity of human activities on epidemic
spreading, this paper studies the susceptible-infected model on a
fully mixed population, where each individual acts in a completely
homogeneous way but different individuals have different mean
activities. Extensive simulations show that the heterogeneity of
activities at population level remarkably affects the speed of
spreading, even though each individual behaves regularly. Further
more, the spreading speed of this model is more sensitive to the
change of system heterogeneity compared with the model consisted of
individuals acting with heavy-tailed inter-event time distribution.
This work refines our understanding of the impact of heterogeneous
human activities on epidemic spreading.

\end{abstract}


\end{frontmatter}

\section{Introduction}

To explain, predict and control the epidemic spreading, studies on
infectious diseases and computer virus attract increasing
attention of many branches of science and engineering, ranging
from mathematics, physics to biology and sociology
\cite{Anderson1991,Bailey1975,Hethcote2000}. Many ingredients
affect the spreading processes, such as the infectivities of
biological virus \cite{Meyers2005,Yang2009} and the
multi-dimensional infection mechanisms of mobile virus
\cite{Wang2009}. Recently, quantitative understanding of human
behavior has refined the traditional models and results
\cite{Barabasi2007,Zhou2008}. These include the structure of human
contact networks, the traveling patterns and the temporal
regularities of human activities. In a word, all of them display
heterogeneous natures, which have remarkable impacts on epidemic
spreading.

The inchoate assumptions are limited to spreading on a homogeneous
network such as ER network or lattices \cite{Gallagher2008} or
ignoring the network structure on the hypothesis that all the
individuals in a system are in the vicinity
\cite{Anderson1991,Bailey1975}. With the small-world
\cite{Watts1998} and scale-free networks \cite{Barabasi1999} put
forward successively, the development of complex networks breathes
new impetus into the study. Increasing empirical data indicate
that, not only in the real social networks, but in the virtual
webs as well, the degrees of individuals can be better
characterized by a heavy-tailed distribution, deviating from the
former Poisson assumption. The hub nodes play a role of
`super-spreaders', having a striking impact on spreading process
\cite{Pastor-Satorras2003,Zhou2006}. It has been demonstrated that
the threshold of spreading on the heterogeneous networks is
remarkably smaller than that of the homogeneous networks
\cite{Barrat2000,Pastor2001}. Recent empirical study indicates
that the displacements of people's traveling have the similar
heterogeneity that short movements are overwhelming majority while
long movements essentially sprinkle \cite{Brockmann2006}. The
spatial structure and human mobility have great effects on the
epidemic spreading
\cite{Gonzalze2004,Gonzalze2006,Lind2007,Tang2009}. Especially,
the heavy-tailed displacement, integrated with the heterogeneity
of networks, fasten the rate of spreading especially in the global
spatial transportation \cite{Ni2009}. In addition to the
heterogeneous character materialized in spatial structure, the
temporal activities, characterized by the inter-event and response
times, have considerable impacts on epidemic spreading and can not
be simply approximated by uniform distributions.

Recent empirical studies indicate that the human activities,
quantified by both the inter-event time and response time, display a
heavy-tailed nature that can not be well characterized by the
Poissonian approximation \cite{Barabasi2007,Zhou2008}. Examples
include the email communication \cite{Barabasi2005}, the cell-phone
communication \cite{Candia2008}, the short-message communication
\cite{Hong2009,Wu2010,Zhao2011}, the web page visits
\cite{Dezso2006,Goncalves2008,Radicchi2009} and some other online
activities \cite{Leskovec2007,Zhou2008b}. To name just a few. The
existence of the power-law-like distribution of inter-event time in
the population level (i.e., for a crowd of people) has already been
well accepted in scientific community, yet the understanding in
individual level is still under debate. Although the mainstream
opinion is that the individuals also have heavy-tailed temporal
activities \cite{Barabasi2005,Vazquez2006}, some scientists, from
both theoretical and experimental aspects, pointed that Poissonian
individuals with different acting rates and periodical activity can
also lead to heavy-tails in the population level
\cite{Hidalgo2006,Malmgren2008}. As we know, some attempts have
already been made to understand the influence of human dynamics,
indicating that the heterogeneity and burstiness of human activities
have striking effects on the speed of spreading
\cite{Iribarren2009,Vazquez2007,Karsai2010}. These works base on a
strong assumption that individuals display heavy-tailed temporal
behavior (which will of course lead to heavy-tailed inter-event time
distribution in the population level). Although it is reasonable for
a number of real systems, it may fail in explaining some other
systems with Poissonian-like individuals and mathematically
speaking, it is still unclear to us whether the heterogeneity in
population level will affect the spreading speed if each individual
is homogeneous.

We try to answer the above question according to a toy model where
each individual behaves in a constant rate but the rates of
different individuals are different, following a power-law
distribution. To keep simplicity, we applied the simplest epidemic
spreading model, the so-called susceptible-infected (SI) model
\cite{Barthelemy2004,Zhou2005,Yan2005,Vazquez2006b,Zhou2006b,Bai2007},
in a fully mixing environment without consideration of the effects
of network structure and spatial locations. Simulation results show
that even though every individual is homogeneous, the heterogeneity
in the population level has remarkable effects on the epidemic
spreading: the larger heterogeneity of individuals' activities
results in the faster spreading. In comparison with another toy
model where individuals are identical to each other but each
individual displays power-law inter-event time distribution, we can,
to some extent, distinguish the effects from population level and
individual level. Our simulation results indicate that the
heterogeneity in the population level has higher impact rather than
that in the individual level, which has refined some previous
understanding.

\begin{figure}
\begin{center}
\includegraphics*[scale=0.35]{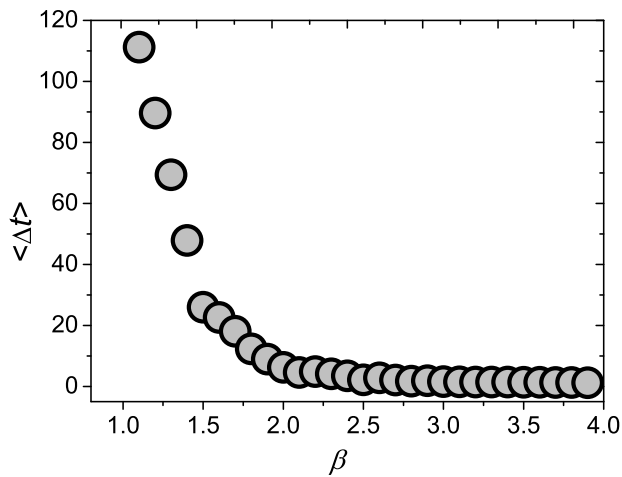}
\end{center}
\caption{The average inter-event time versus power-law exponent.
Each circle corresponds to an average value over $10^5$ data
points and the value of $M$ is set as 1000.}
\end{figure}

\section{Model}

In the SI model, each individual can only be in two discrete
states, either susceptible or infected. This model is usually
applied to describe early epidemic stages in which no control
measures are deployed, and after infected, individuals remain
always infective, and cannot go back to the susceptible state. In
a fully mixing environment, there is a certain possibility for
every infected individual to infect all the susceptible ones at
each time step. Speak in images, the infected individual can be
associated with gaseous molecules, who can move with freedom to
infect all of other individuals in certain probability. In this
model, the infectivity rate $\lambda$, at which susceptible
individuals acquire the infection from an infected individuals,
and the total population $N$ are set as constants.

The susceptible agents are the same to the ones in the traditional
model, namely at each time step they are probably infected if some
infected agents contact them. In contrast, at a given time step, an
infected agent can be inactive, that is to say, she/he will not
contact any other susceptible agents at that time step, while if
she/he is active, she/he will contact every susceptible agents with
infectivity rate $\lambda$. The inter-event time is then defined as
the time difference between two consequent active steps of an
infected individual. Each infected individual acts in an identical
rate with an unchanged inter-event time. Denote by $\Delta t_{i}$
the inter-event time of the $i$th individual, $i$ will be active at
the time steps $t_{i0}, t_{i0}+\Delta t_{i}, t_{i0}+2\Delta t_{i},
\cdots$, where $t_{i0}$ is the first active time step for $i$. To
avoid the bias caused by synchronized actions, for each individual
$i$, the starting time $t_{i0}$ is randomly selected in the interval
$[0, \Delta t_{i}-1]$. Through keeping inter-event time of an
individual a constant, we can get rid of the influence brought by
the temporal heterogeneity in the individual level and concentrate
on the impact from the heterogeneity of the inter-event times of
different individuals. We assume that the distribution of $\Delta t$
obeys a power-law form as $P(\Delta t)\sim \Delta t^{-\beta}$.
Initially, each individual samples an inter-event time from this
distribution.

\begin{figure}
\begin{center}
\includegraphics*[scale=0.85]{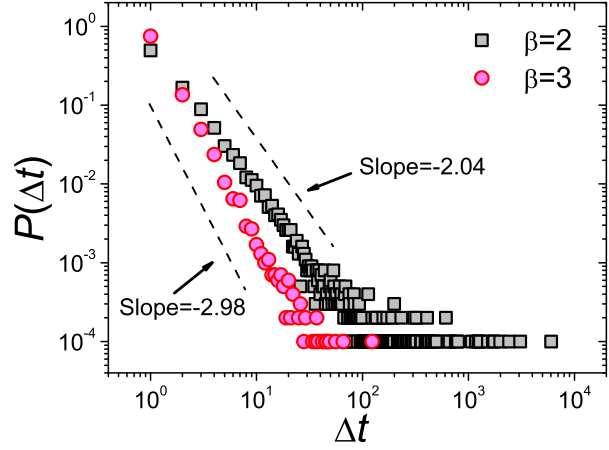}
\end{center}
\caption{Two typical examples of distributions of generated
inter-event times. The black squares and red circles denote the
cases of $\beta=2$ and $\beta=3$, respectively. The number of
individuals is $N=1000$ and the cell limit is $M=1000$. In this
figure, we have not considered the normalization by $m(\beta)$.}
\end{figure}

Power-law distributions come in two basic favors: continuous
distributions govern continuous real numbers and discrete
distributions consider integers. The latter form is needed here,
but the way to produce it requires us think about the continuous
case. Avoiding some strikingly long time steps, we set a cell
limit $M$, and the task turns to be the generating of the
following probability function
\begin{equation}
P(k)\propto k^{-\beta},
\end{equation}
where $k$ is an integer in the range $[1,M]$. Accordingly, the
cumulative distribution reads
\begin{equation}
F(k)=\sum_{1}^{k} P(k),
\end{equation}
and can be approximated by a continuous form as:
\begin{equation}
F(k)=\int_{1}^{k} P(k).
\end{equation}

Denote by $r$ the ratio of $F(\Delta t)$ ($1\leq \Delta t \leq M$)
to $F(M)$, as
\begin{equation}
r=\frac{\int_{1}^{\Delta t} x^{-\beta} dx}{\int_{1}^{M} x^{-\beta}
dx}=\frac{\frac{\Delta
t^{1-\beta}-1}{1-\beta}}{\frac{M^{1-\beta}-1}{1-\beta}}=\frac{\Delta
t^{1-\beta}-1}{M^{1-\beta}-1}.
\end{equation}
Clearly, $0\leq r \leq 1$ and $\Delta t$ can be obtained by:
\begin{equation}
\Delta t=(r\times(M^{1-\beta}-1)+1)^{\frac{1}{1-\beta}}.
\end{equation}
In the large limit of $M$, $M^{1-\beta}$ is close to zero and thus
Eq. 5 can be approximated as
\begin{equation}
\Delta t \approx (1-r)^{\frac{1}{1-\beta}},
\end{equation}
which is in accordance with the result shown by Clauset, Shalizi
and Newman \cite{Clauset2009}.

A serious problem in the above method is that for different
$\beta$, the average inter-event times $\langle \Delta t\rangle$
are different, and thus it is not fair to compare systems with
different $\beta$. In fact, when $\beta\leq 2$, the average
inter-event time diverges for the infinite large systems. Figure 1
reports a simulation result for a finite system, from which it is
obvious that when $\beta$ decreases from 2 to 1, the average
inter-event time grows very fast. To eliminate the biased
influence caused by different $\langle \Delta t \rangle$, we
introduce a factor $m(\beta)$ to Eq. 5, as
\begin{equation}
\Delta
t'=m(\beta)\times(r\times(M^{1-\beta}-1)+1)^{\frac{1}{1-\beta}},
\end{equation}
where $m(\beta)$ satisfies
\begin{equation}
\frac{\langle \Delta t\rangle_{\beta_1}}{\langle \Delta
t\rangle_{\beta_2}}=\frac{m(\beta_{2})}{m(\beta_{1})}.
\end{equation}
Without the loss of generality, we set $m(1)=1$ and thus
\begin{equation}
m(\beta)=\frac{\langle\Delta t\rangle_{1}}{\langle\Delta
t\rangle_\beta}.
\end{equation}

To summarize, in the initialization, given $M$ and $\beta$,
$m(\beta)$ can be estimated directly by simulation, and then for
each individual $i$, we first generate a random real $r$ in the
range $[0,1]$ and then calculate her/his characteristic
inter-event time $\Delta t_i$ according to Eq. 7.

\begin{figure}
\begin{center}
\includegraphics*[scale=0.2]{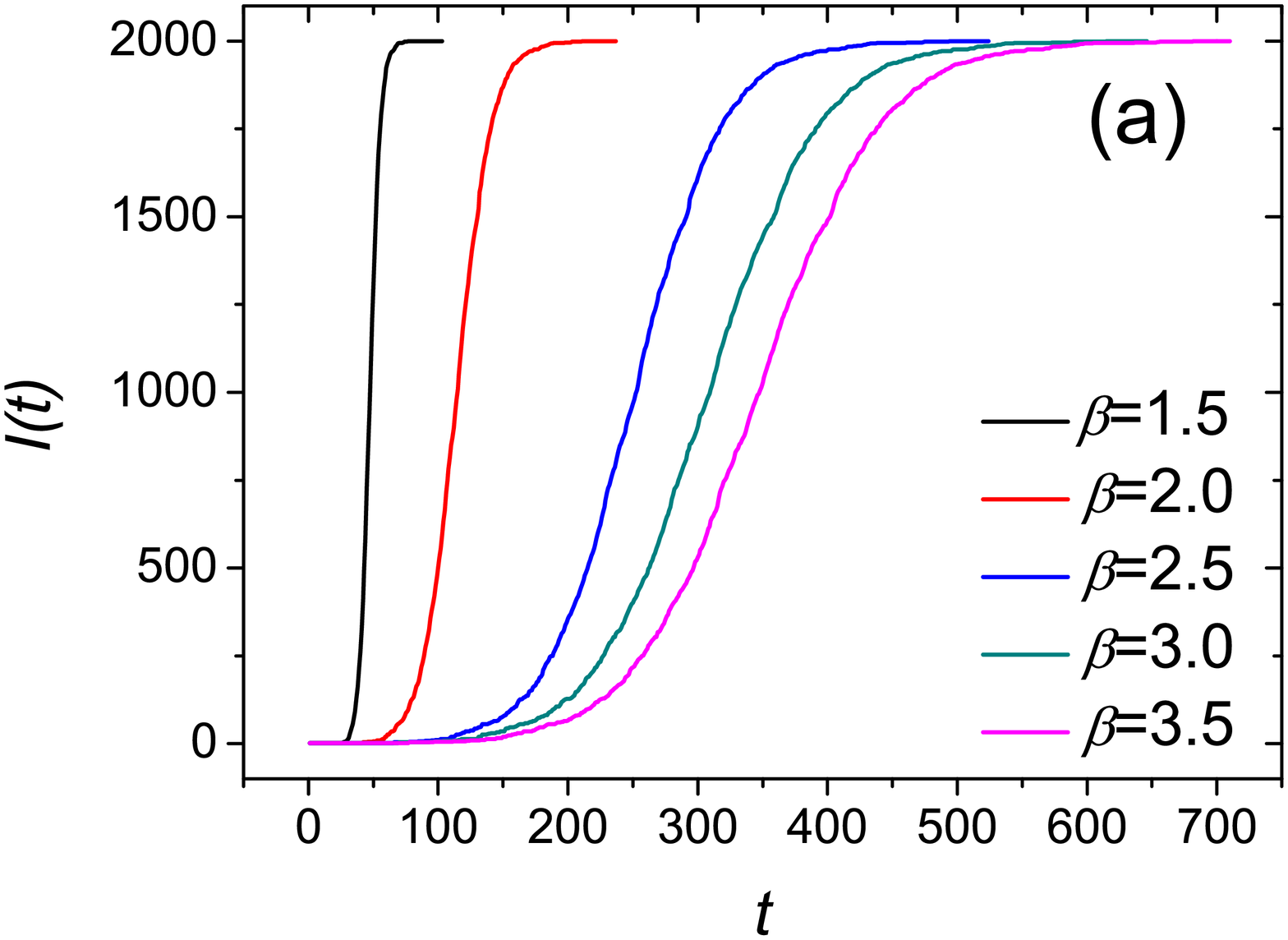}
\includegraphics*[scale=0.2]{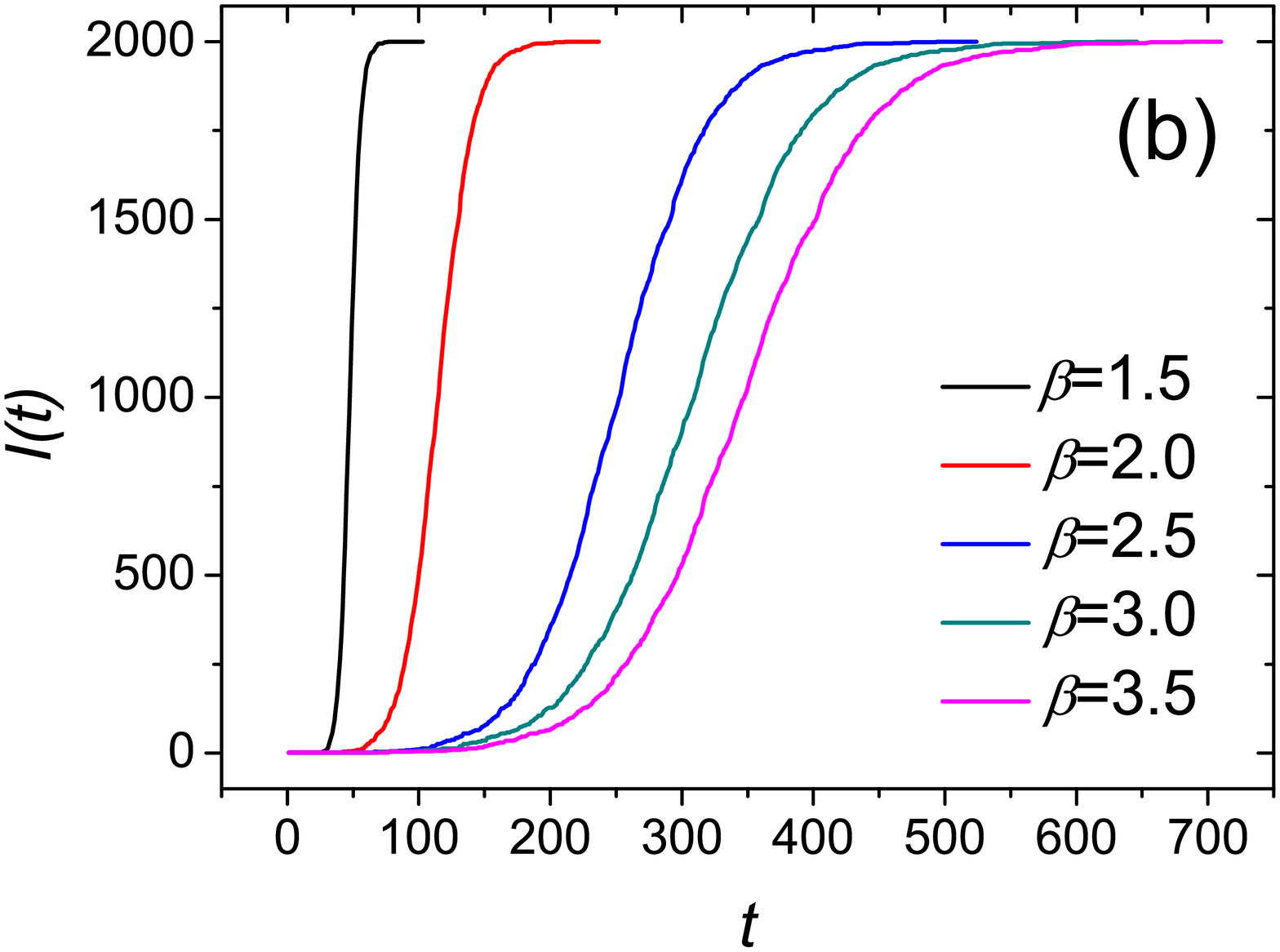}
\includegraphics*[scale=0.2]{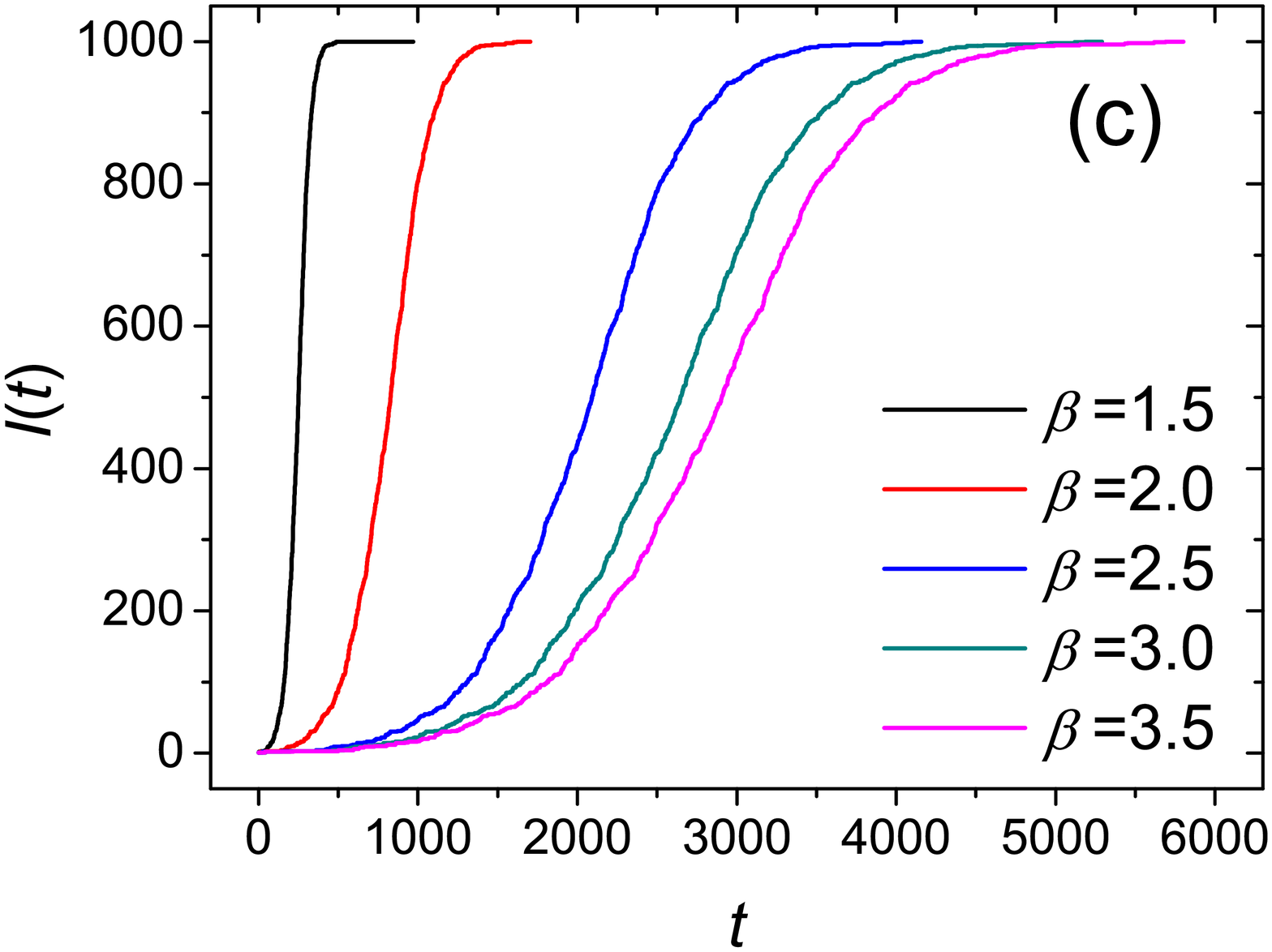}
\scalebox{0.2}[0.183]{\includegraphics{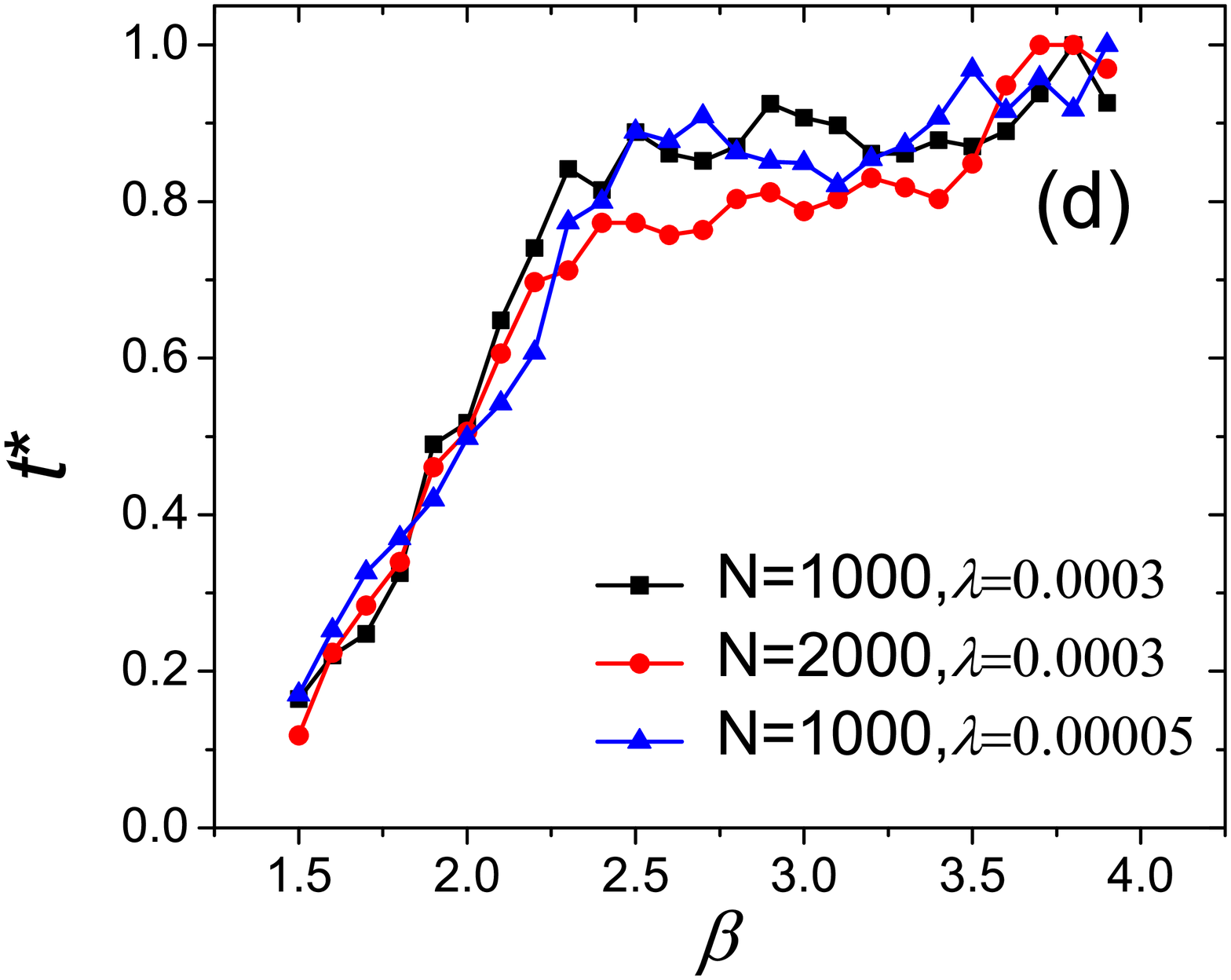}}
\end{center}
\caption{Impacts of the power-law exponent $\beta$ on the epidemic
spreading process. The first three plots show the number of
infected nodes $I(t)$ versus the time step $t$, where the
parameters are: (a) $N=1000$, $\lambda=3\times 10^{-4}$, (b)
$N=2000$, $\lambda=3\times 10^{-4}$ and (c) $N=1000$,
$\lambda=5\times 10^{-5}$. The plot (d) displays the first time
step $t^*$ when the number of infected nodes exceeds half of the
population. In plot (d), black squares, red circles and blue
triangles correspond to the cases shown in plots (a), (b) and (c),
respectively. All the data points are obtained by averaging over
10 independent runs.}
\end{figure}

Notice that, the generated inter-event time from Eq. 7 may be not
well-pleasing for our model since it is generally not an integer.
Here, for a real $\Delta t$, we firstly separate its integral and
decimal parts as $\Delta t=\lfloor\Delta t\rfloor+b$, $b\in[0,1)$.
Then we reset $\Delta t \leftarrow \lfloor\Delta t\rfloor$ with
probability $1-b$, while $\Delta t \leftarrow \lfloor\Delta
t\rfloor+1$ with probability $b$. For example, if we obtain
$\Delta t=3.2$ from Eq. 7, it will be reset as 3 with probability
0.8 while 4 with probability 0.2. Figure 2 shows typical
distributions for a system with $N=1000$ individuals for different
$\beta$, from which one can see that the generating method for
heterogeneous inter-event times well meets the requirement (The
fitting exponents, shown as slopes in the figure, are obtained by
using the maximum likelihood method \cite{Goldstein2004}).

\section{Results and Discussion}

A set of comparative simulations are carried out to show the
impacts of heterogeneous activity on spreading speed. Each run
starts with one randomly selected node as infected node and the
other $N-1$ nodes all susceptible. Figure 3 reports how the
spreading speed, characterized by the number of infected nodes
$I(t)$, is affected by the heterogeneity of the individual
activities. Although each individual behaves in a perfectly
uniform pace, the heterogeneity at the population level has
remarkably impacts on the epidemic spreading, and the spreading
can be accelerated by enhancing the heterogeneity (i.e., reducing
the exponent $\beta$). To clearly show the relation between speed
and heterogeneity, we further consider the required time steps
$t^*$ to infect half population from the initialization. We
normalize $t^*$ by dividing by $t^*_{\texttt{max}}$ given a
parameter set $(N,\lambda)$, which is helpful in better displaying
several $t^*(\beta)$ curves with different parameter sets
$(N,\lambda)$. One can see from figure 3(d), $t^*$ monotonously
increases with $\beta$, again indicating that the spreading can be
accelerated by reducing $\beta$.

\begin{figure}
\begin{center}
\includegraphics*[scale=0.2]{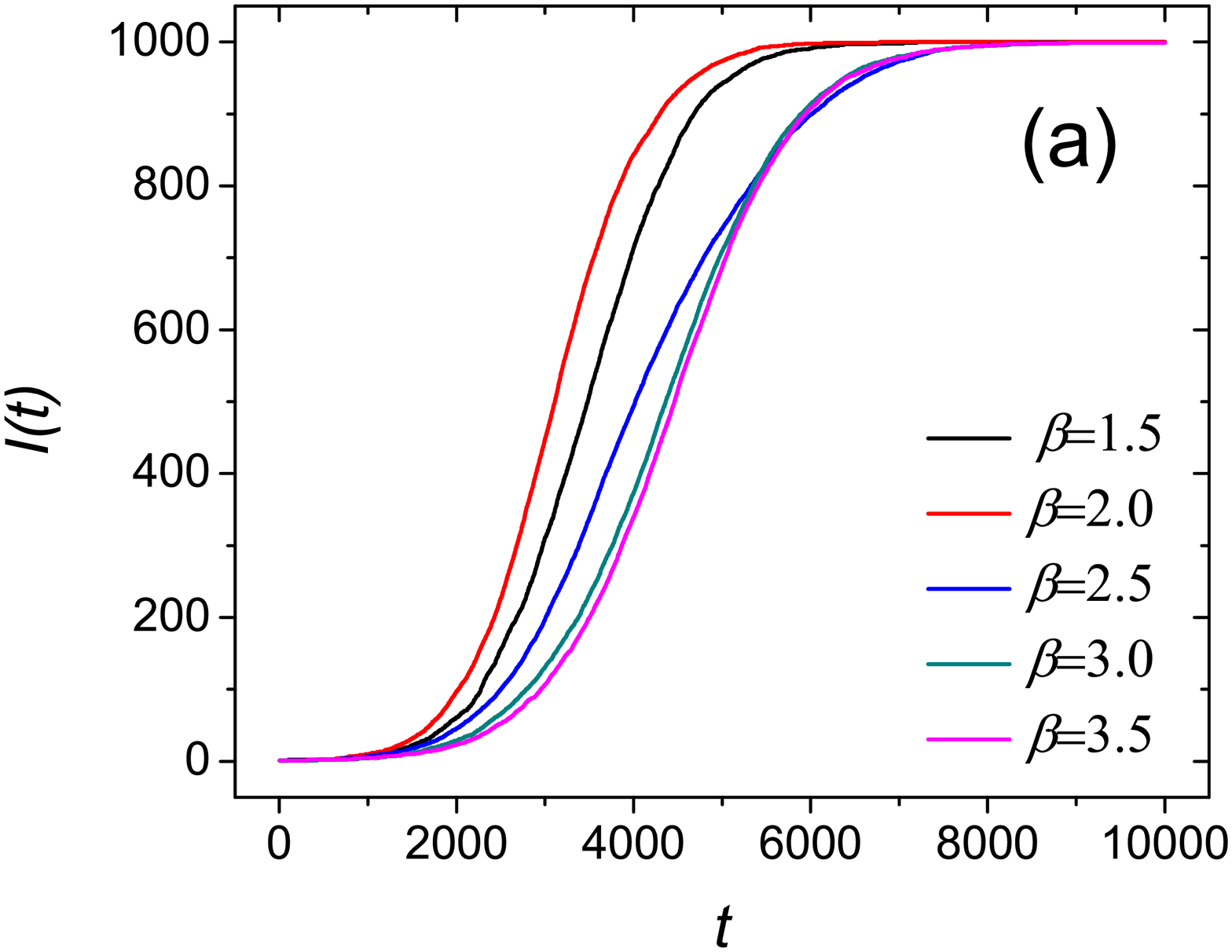}
\includegraphics*[scale=0.2]{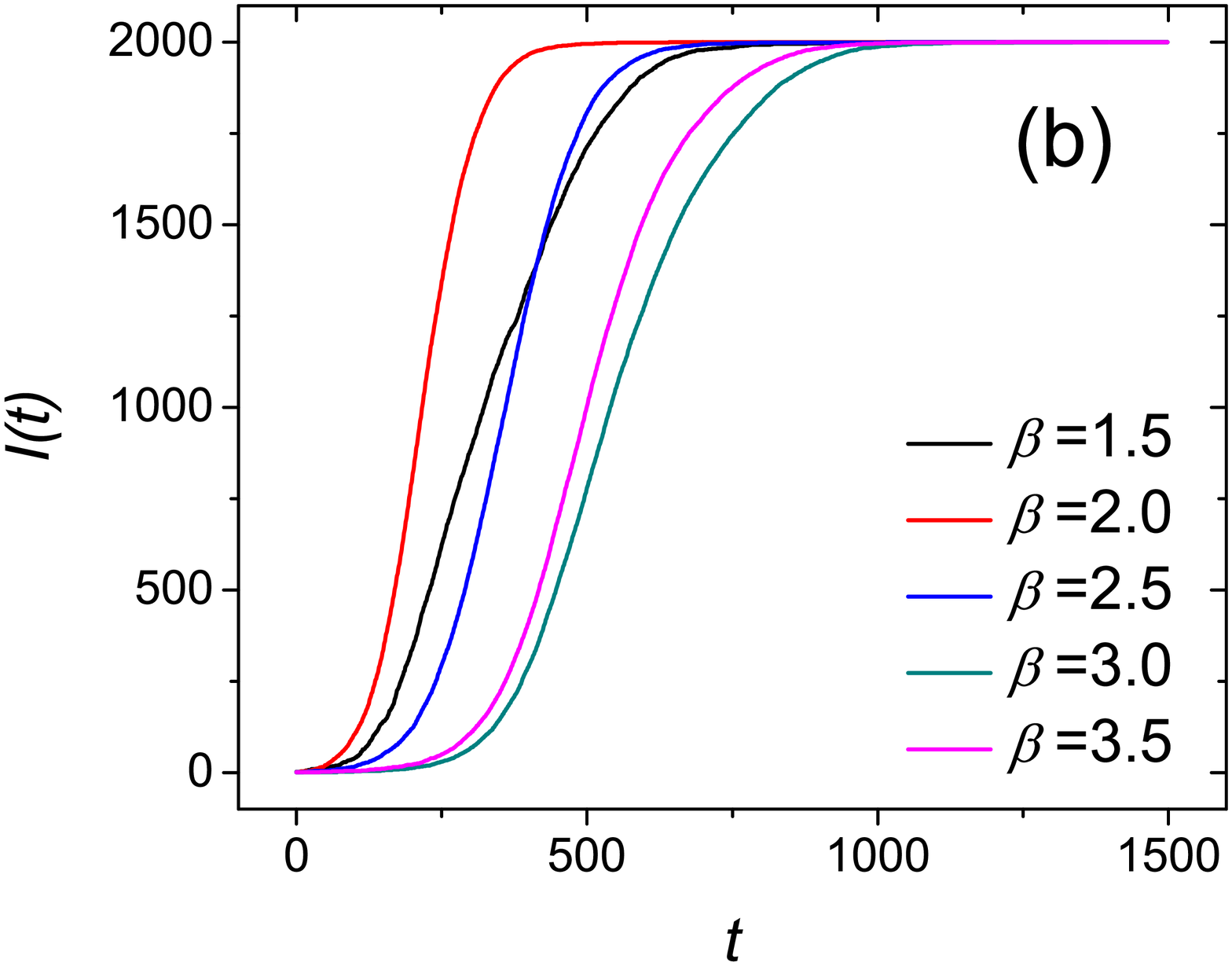}
\includegraphics*[scale=0.2]{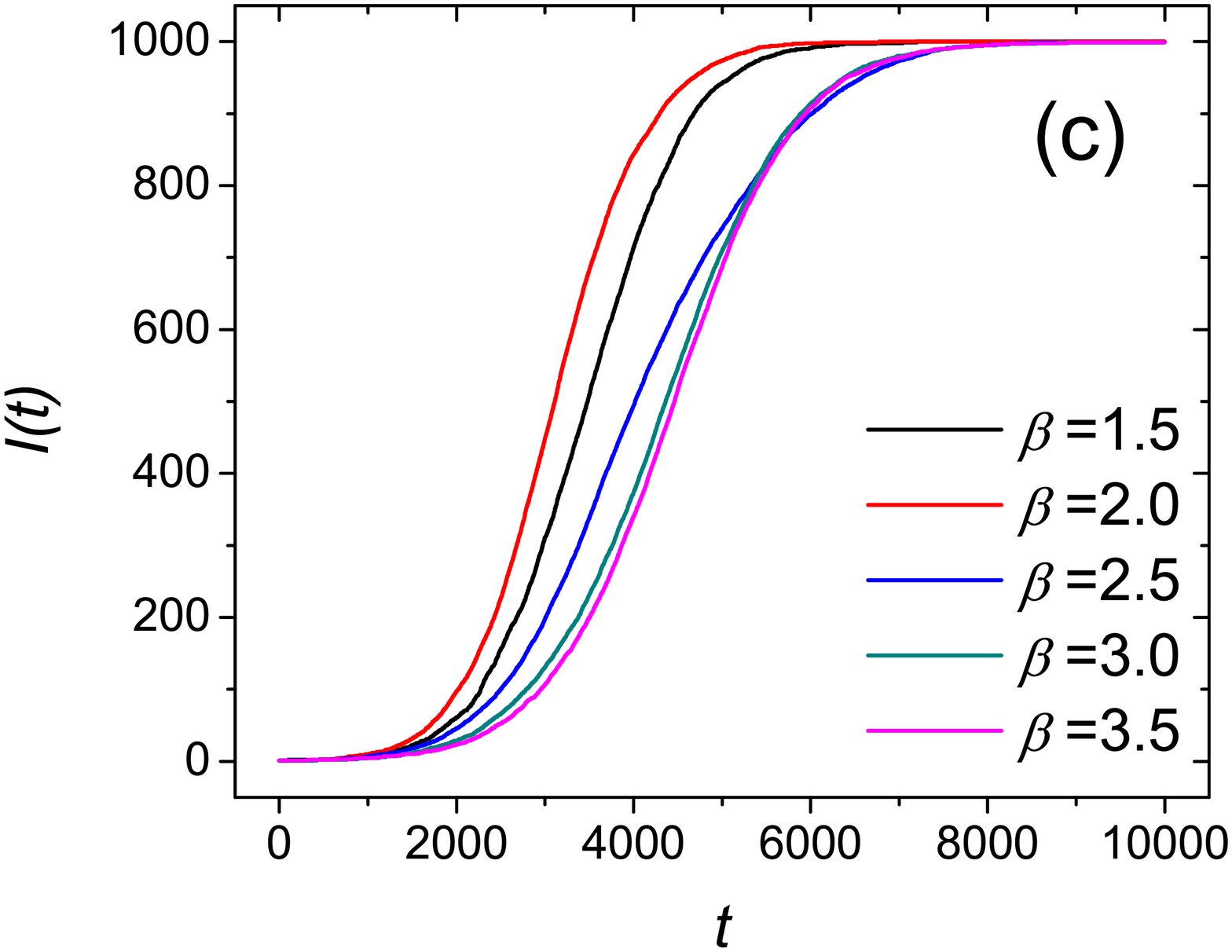}
\includegraphics*[scale=0.2]{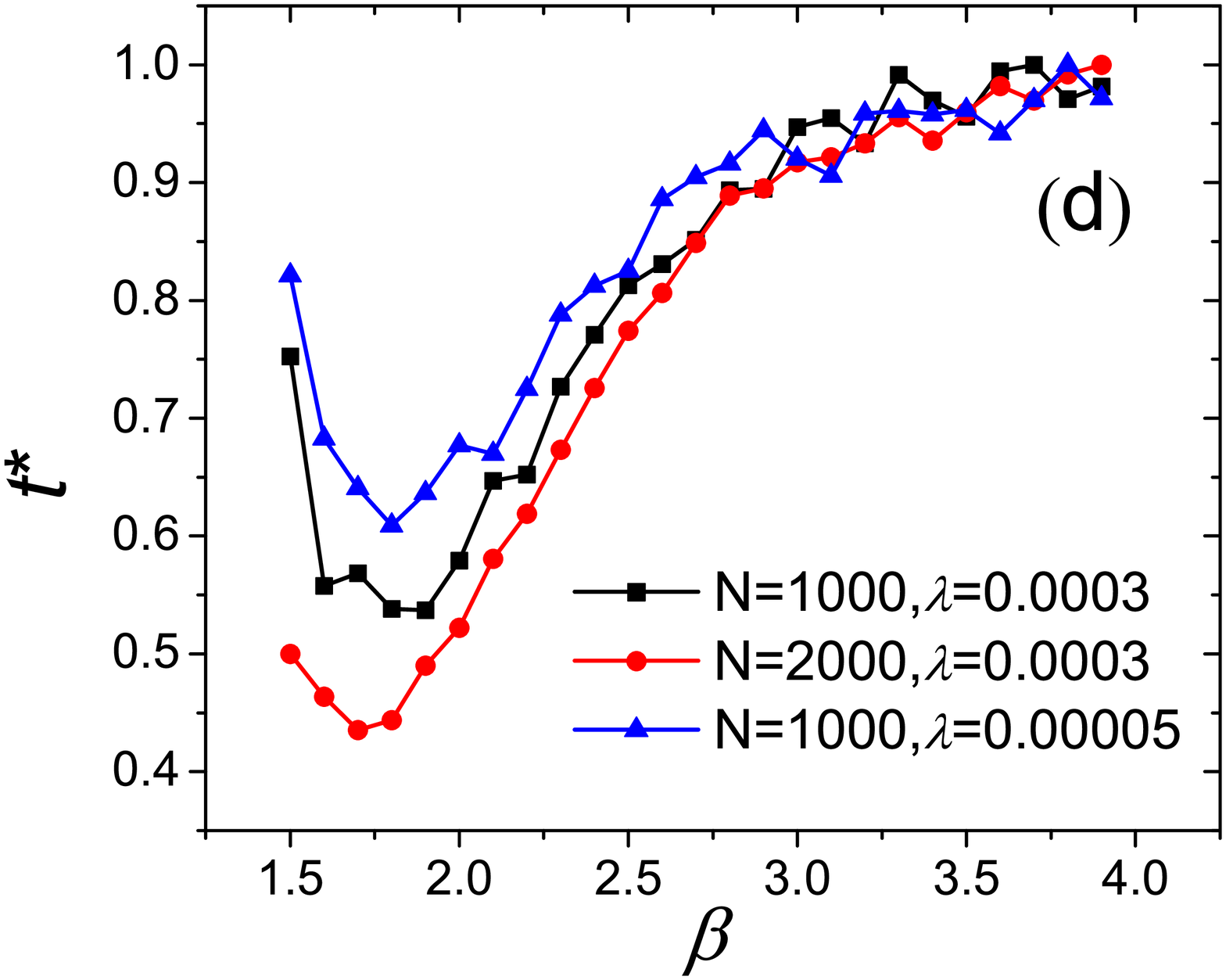}
\end{center}
\caption{Impacts of the power-law exponent $\beta$ on the epidemic
spreading process of the model for comparison. The first three
plots show the number of infected nodes $I(t)$ versus the time
step $t$, where the parameters are: (a) $N=1000$, $\lambda=3\times
10^{-4}$, (b) $N=2000$, $\lambda=3\times 10^{-4}$ and (c)
$N=1000$, $\lambda=5\times 10^{-5}$. The plot (d) displays the
first time step $t^*$ when the number of infected nodes exceeds
half of the population. In plot (d), black squares, red circles
and blue triangles correspond to the cases shown in plots (a), (b)
and (c), respectively. All the data points are obtained by
averaging over 100 independent runs.}
\end{figure}

As we have mentioned before, the heavy-tailed distribution of
inter-event time at the population level is well accepted, but
whether the individual is heterogeneous is still under debate
\cite{Hidalgo2006,Malmgren2008}. With the assumption that each
individual has heavy-tailed temporal activities, previous works
have revealed the non-ignorable impacts on epidemic spreading
\cite{Iribarren2009,Vazquez2007,Karsai2010}, yet they have not
answered the question whether the heterogeneity in population
level will affect the spreading speed if each individual is
homogeneous. Figure 3 clearly says ``YES"!

To make a comparison, we next investigate another model where each
individual acts with a power-law inter-event time distribution
$P(\Delta t)\sim \Delta t^{-\beta}$. Here $\beta$ is a system
parameter, namely the inter-event time distributions of all
individuals are the same. Of course, at the population level, the
inter-event time distribution is also $P(\Delta t)\sim \Delta
t^{-\beta}$. This model is closer to the assumption of previous
works \cite{Iribarren2009,Vazquez2007}. As shown in Fig. 4, the
exponent $\beta$ also affects the spreading process, but in a more
complicated way \cite{ex1}. In the range $\beta\in [1.5, 4]$ and for
$N=2000$, subject to $t^*$, the fastest spreading is about twofold
faster than the slowest one in Fig. 4(d), while in Fig. 3(d), it can
be tenfold faster. In a word, whatever the individual activities are
homogeneous or not, the heterogeneity at the population level has
remarkable impact on epidemic spreading. In comparison, a system
consisted of heterogeneous individuals is more sensitive to the
temporal heterogeneity of activities at the population level even
though each individual acts in a completely periodical way.
Qualitatively speaking, the difference between a system of regular
agents with different acting rates and a system of non-Poissonian
agents with the same statistics is analogous to the difference
between quenched and annealed systems. In the former system, a few
agents are born with very large $\Delta t$, which play the role of
absorbing nodes that are hardly to infect other agents. In the
latter system, each agent has the chance to suffer a very large
$\Delta t$, but not usual. These two systems can be considered as a
quenched system and an annealed system, respectively. We believe
that the difference of impacts reported in this paper is similar to
the difference of spreading processes in quenched and annealed
networks \cite{Doussal1989,Baronchelli2010}, however, to which
extent the observed different impacts can be explained by the
difference between quenched and annealed systems, as well as how to
characterize and understand the difference between quenched and
annealed systems are still open questions to us. To summarize, this
work provides complementary information to the previous studies and
refines our understanding of the impact of heterogeneous human
activities on epidemic spreading.

\section*{Acknowledgments}

We acknowledge Duan-Bing Chen, Ming-Sheng Shang, Ming Tang and
Zhi-Dan Zhao for their valuable discussions. This work is
partially supported by the Yellow Ginkgo Project from the
University of Electronic Science and Technology of China, and the
National Natural Science Foundation of China under Grant Nos.
70871082, 10975126 and 70971089.


\begin{thebibliography}{0}
\bibitem{ex1} It is very interesting that the effetc of $\beta$ is
not ``monotonous". As shown in Fig. 4(a)-4(c), the process with
$\beta=2.0$ is the fastest, and more clearly can be found in Fig.
4(d), $\beta$ has an ``optimal" value subject to the fastest
spreading at about 1.8. This complicated phenomenon has not been
observed in the previous models and we will discuss it (with
analytical solution on $I(t)$) in another paper.
\end{thebibliography}
\end{document}